\documentclass[aip,reprint]{revtex4-1}
\usepackage{amsmath,empheq}
\usepackage{amssymb}
\usepackage{amsfonts}
\usepackage{float}
\usepackage[T1]{fontenc}
\usepackage{bm}
\usepackage{color}
\usepackage{graphicx}
\usepackage{hyperref}
\hypersetup{
    bookmarks=true,         
    unicode=false,          
    pdftoolbar=true,        
    pdfmenubar=true,        
    pdffitwindow=false,     
    pdfstartview={FitH},    
    pdftitle={Conservative discretization of the Landau collision integral},    
    pdfauthor={Eero Hirvijoki},     
    pdfsubject={Discretization of the Landau operator},   
    pdfcreator={pdflatex},   
    pdfproducer={pdflatex}, 
    pdfkeywords={collision operator} {conservation laws}, 
    pdfnewwindow=true,      
    colorlinks=true,        
    linkcolor=black,      
    citecolor=blue,         
    filecolor=blue,      
    urlcolor=blue           
}
\usepackage{array}
\usepackage[caption=false]{subfig}

\newcolumntype{C}[1]{>{\centering\arraybackslash$}p{#1}<{$}}

\makeatother

\begin{document}

\title{Conservative discretization of the Landau collision integral}

\author{E. Hirvijoki}
\email{ehirvijo@pppl.gov}
\affiliation{Princeton Plasma Physics Laboratory, Princeton, NJ, USA}
\author{M. F. Adams}
\email{mfadams@lbl.gov}
\affiliation{Lawrence Berkeley National Laboratory, Berkeley, CA, USA}

\date{\today}

\begin{abstract}
  We describe a density-, momentum-, and energy-conserving
  discretization of the nonlinear Landau collision integral. The
  method is suitable for both the finite-element and discontinuous
  Galerkin methods and does not require structured meshes. The
  conservation laws for the discretization are proven algebraically
  and demonstrated numerically for an axially symmetric nonlinear
  relaxation problem using a finite-element implementation.
\end{abstract}


\maketitle

\section{Introduction}
Structure preserving numerical methods, which preserve a property or
{\it structure} of the equation, are becoming not only important but
necessary. For instance, it is widely recognized that integration of
Hamiltonian systems with non-structure-preserving methods leads to
numerical dissipation that may render the solution useless. This is
especially true if the numerical method is to track the long
time-scale behaviour of the system, as often is the case.

For many of the purely Hamiltonian or variational systems encountered
in plasma physics, recent research has provided structure-preserving
discretization
methods~\cite{Qin_et_al_2008,Squire_et_al_2012,Kraus2013thesis,
  lelandellison:ppcf:2015,Qin_et_al_2016_nf,He_et_al_2016,Zhou_et_al_2014}. Yet
dissipative effects reside on a largely uncharted territory: the
structure, as it is understood for Hamiltonian or variational systems,
is not that well understood for general dissipative
systems. Exceptions do exist and, in the case of Coulomb collisions,
the Landau collision integral~\cite{Landau_1937} can, in fact, be
described in terms of a {\it symmetric} bracket~\cite{Morrison_1986}
(whereas Hamiltonian dynamics emerges from an antisymmetric Poisson
bracket). Unfortunately, it is not clear yet how the symmetric, {\it
  metric} bracket should be discretized to preserve the underlying
structural properties of the Landau collision integral.

Until an appropriate discretization of the so-called {\it
  metriplectic} formulation, describing both the Hamiltonian and
dissipative dynamics, is discovered, one could consider the discrete
version of the {\it Lagrange-d'Alembert principle} and embed a
discrete Landau collision integral into structure preserving
discretizations of the Vlasov-Maxwell system. As finite-element and
discontinuous Galerkin methods are receiving increasing attention for
addressing the Vlasov-Maxwell part of the kinetic system, we find it
appealing to study how these methods would adapt to addressing the
Landau collision integral.

The result we provide in this paper is a proof that discretization of
the nonlinear Landau collision integral, with a set of basis functions
capable of presenting second-order polynomials exactly, guarantees
exact discrete conservation laws for density, momentum, and
energy. Effectively, this observation opens up the possibility to use,
e.g., the adaptive finite-element or discontinuous Galerkin technology
to efficiently capture regions of sharp gradients in the velocity
space while still preserving the underlying structure in the numerical
solution.

As we shall show, the only requirement is to retain the Landau's
original integral formulation and not resort to the
Rosenbluth-MacDonald-Judd potential formulation~\cite{RMJ1957}: while
the potential formulation is efficient in decreasing the numerical
burden of evaluating the collision integral from ${\cal O}(N^2)$ down
to ${\cal O}(N \log N)$, it simultaneously destroys numerical
conservation laws so that artificial modification of the collision
integral is necessary~\cite{Taitano_2015}. We also note that the
${\cal O}(N^2)$ part of our method belongs to the class of so-called
{\it embarassingly parallel problems} and it is thus expected to scale
well on highly vectorized platforms. Indeed, the feasibility of a
different ${\cal O}(N^2)$-algorithm has already been
demonstrated~\cite{Yoon_et_al_2014,Hager_2016}.

For the sake of compactness, our proof is provided for a
single-species plasma but it is straightforward to generalize the
proof to handle multi-species plasmas with arbitrary mass and charge
ratios. Also, our discussion concerns the particle phase-space
collision operator, and the axially symmetric numerical example that
is provided is not to be mistaken as a proof for a gyrokinetic
collision operator. Strcture-preserving discretization of the proper
gyrokinetic collision operator~\cite{gyrokinetic-operator} is a far
more challenging issue and is not addressed in the current paper.

The rest of the paper is following: The Landau collision integral and
its properties are reviewed in
Section~\ref{sec:collision_integral}. The discretization, together
with an algebraic proof of the related conservation laws, is detailed
in Section~\ref{sec:discretization}. In
Section~\ref{sec:axial_symmetry}, the claimed properties are
demonstrated numerically for an axially symmetric relaxation problem,
and a summary of the results is given in Section~\ref{sec:summary}.

\section{The Landau collision integral}
\label{sec:collision_integral}
For clarity, we consider the like-species collisions, while the
results generalize for multi-species collisions as well. Next, we
review the explicit form of the collision integral, normalize it to
dimensionless variables, and discuss the collisional invariants.

\subsection{Single species collision operator}
Under small-angle dominated Coulomb collisions, the evolution of the
distribution function $f(t,\bm{u})$ in velocity space
$\bm{u}\in\mathbb{R}^3$ is determined by the integro-differential
equation~\cite{Landau_1937}
\begin{equation}
  \frac{\partial f}{\partial t}= \nu\frac{\partial}{\partial\bm{u}}\cdot\int_{\mathbb{R}^3} d\bm{\bar{u}}\;\mathbf{U}(\bm{u},\bm{\bar{u}})\cdot\left(\bar{f}\frac{\partial f}{\partial\bm{u}}-f\frac{\partial\bar{f}}{\partial\bm{\bar{u}}}\right).
\end{equation}
Here $\nu=e^4\ln\Lambda/(8\pi m^2\varepsilon_0^2)$ can be considered a
reference collision frequency, $\ln\Lambda$ is the Coulomb logarithm,
$e$ and $m$ are the charge and mass, $\varepsilon_0$ is the
  vacuum permittivity, and $\bm{u}=\bm{p}/m$ is the
momentum-per-rest-mass. The quantities with an overbar are evaluated
at $\bm{\bar{u}}$.

The Landau tensor $\mathbf{U}(\bm{u},\bm{\bar{u}})$, valid at
non-relativistic energies, is a scaled projection matrix of the
relative velocity $\bm{u}-\bm{\bar{u}}$ between the colliding
particles:
\begin{equation}
\label{eq:landau_tensor}
\mathbf{U}=\frac{1}{\lvert\bm{u}-\bm{\bar{u}}\rvert^3}\left(\lvert\bm{u}-\bm{\bar{u}}\rvert^2\mathbf{I}-(\bm{u}-\bm{\bar{u}})(\bm{u}-\bm{\bar{u}})\right).
\end{equation}
In the relativistic case, the correct expression for the tensor
$\mathbf{U}(\bm{u},\bm{\bar{u}})$ was derived by Beliaev and
Budker~\cite{Beliaev_Budker_1956}
\begin{equation}
\label{eq:beliaev_budker_tensor}
\mathbf{U}_{\textrm{BB}}=\frac{r^2}{\bar{\gamma}\gamma w^3}\left(w^2\mathbf{I}-\bm{u}\bm{u}-\bm{\bar{u}}\bm{\bar{u}}+r(\bm{u}\bm{\bar{u}}+\bm{\bar{u}}\bm{u})\right),
\end{equation}
where $r=\gamma\bar{\gamma}-\bm{u}\cdot\bm{\bar{u}}/c^2$,
$w=c\sqrt{r^2-1}$, $\gamma=\sqrt{1+u^2/c^2}$,
$\bar{\gamma}=\sqrt{1+\bar{u}^2/c^2}$, and $c$ is the speed of
light. In the limit $c\rightarrow \infty$, the Beliaev-Budker tensor
$\mathbf{U}_{\textrm{BB}}$ reduces to Landau tensor and the
relativistic momenta, normalized to the rest mass, reduce to the
non-relativistic expressions for velocities.

Although the focus of this paper is the nonrelativistic limit, we will
show that standard finite-element or discontinuous Galerkin
discretization of the relativistic collision integral will lead to
exact density and momentum conservation while exact energy
conservation would require development of a completely new set of
basis functions.

\subsection{Normalization}
In numerical applications, one should always work in dimensionless
variables to prevent accumulation of floating point errors. This is
achieved by defining $\bm{x}=\bm{u}/c$ and
$\bm{\bar{x}}=\bm{\bar{u}}/c$ with $c$ some positive constant denoting
a reference velocity. In the relativistic case, $c$ would naturally
denote the speed of light but, for the nonrelativistic case, it can be
considered arbitrary, e.g., the thermal velocity. Obviously, $\bm{x}$
and $\bm{\bar{x}}$ are not to be misunderstood as the configuration
space variables.

The velocity-space gradients and differential volume elements
transfrom according to
\begin{align}
&\partial/\partial\bm{u}=c^{-1}\nabla, &\partial/\partial\bm{\bar{u}}=c^{-1}\bar{\nabla},\\
&d\bm{u}=c^3d\bm{x}, &d\bm{\bar{u}}=c^3d\bm{\bar{x}},
\end{align}
while the tensor $\mathbf{U}$ transforms according to
\begin{equation}
\mathbf{U}(\bm{u},\bm{\bar{u}})=c^{-1}\mathbf{U}(\bm{x},\bm{\bar{x}}).
\end{equation}
Further, we normalize time according to
\begin{equation}
t=\frac{8\pi\varepsilon_0^2m^2}{e^4\ln\Lambda}\tau=\nu^{-1}\tau,
\end{equation}
so that the normalized Landau integral equation becomes
\begin{equation}
\label{eq:landau}
\frac{\partial f}{\partial\tau}=\nabla\cdot\int_{\mathbb{R}^3} d\bm{\bar{x}}\;\mathbf{U}(\bm{x},\bm{\bar{x}})\cdot\left(\bar{f}\nabla f-f\bar{\nabla}\bar{f}\right).
\end{equation}

\subsection{Conservation laws}
Without loss of generality, we define a domain $\Omega$ and require
that $f$ vanishes at the boundary $\partial\Omega$. Additionally, the
normal component of the velocity-space flux
\begin{equation}\label{eq:flux}
\bm{J}(\bm{x})\equiv\int_{\Omega} d\bm{\bar{x}}\;\mathbf{U}(\bm{x},\bm{\bar{x}})\cdot\left(\bar{f}\nabla f-f\bar{\nabla}\bar{f}\right),
\end{equation}
is required to vanish at the boundary $\partial\Omega$, to satisfy
density conservation. Obviously, both these conditions are true if
$\Omega$ is chosen to be $\mathbb{R}^3$. In numerical implementations
the domain $\Omega$ must, however, be finite and, thus, we use
$\Omega$ as an arbitrary domain for now.

If we now multiply Eq.(\ref{eq:landau}) with a function $\phi$, and
integrate over the domain $\Omega$ and apply the boundary conditions,
we find
\begin{equation}
  \int d\bm{x}\;\phi\;\frac{\partial f}{\partial \tau}=\int_{\Omega} d\bm{x}\;\nabla \phi\cdot \int_{\Omega}d\bm{\bar{x}}\;\mathbf{U}\cdot\left(f\bar{\nabla}\bar{f}-\bar{f}\nabla f\right).
\end{equation}
Upon rearranging the integration order, one obtains
\begin{equation}
  \int d\bm{x}\;\phi\;\frac{\partial f}{\partial \tau}=\frac{1}{2}\int_{\Omega} d\bm{x}\int_{\Omega}d\bm{\bar{x}}\left(\nabla \phi-\bar{\nabla}\bar{\phi}\right)\cdot\mathbf{U}\cdot\left(f\bar{\nabla}\bar{f}-\bar{f}\nabla f\right).
\end{equation}
For $\phi(\bm{x}) \in \{1,\bm{x}\}$, the above expression obviously
vanishes. Further, $(\nabla {\cal E}(\bm{x})-\bar{\nabla}{\cal
  E}(\bm{\bar{x}}))\cdot\mathbf{U}(\bm{x},\bm{\bar{x}})$ vanishes for
both the non-relativistic energy ${\cal E}(\bm{x})=x^2$ (with
$\mathbf{U}$ the Landau tensor) and the relativistic energy ${\cal
  E}(\bm{x})=\sqrt{1+x^2}$ (with $\mathbf{U}$ the Beliaev-Budker
tensor $\mathbf{U}_{\textrm{BB}}$), due to the properties of the
tensor $\mathbf{U}$ ($\mathbf{U}_{\textrm{BB}}$). Thus, the quantities
$\int_{\Omega} d\bm{x}\; \phi\; f $ are referred to as {\it
  collisional invariants}, whenever $\phi(\bm{x}) \in \{1,\bm{x},{\cal
  E}(\bm{x})\}$.

\section{Discretization}
\label{sec:discretization}
One of the challenges in discretizing the Landau operator is to
preserve the collisional invariants that exist for the continuous
operator. Here we prove that discretization of the weak formulation
\begin{align}
\label{eq:normalized-weak-form}
\int_{\Omega} d\bm{x}\;\phi\;\frac{\partial f}{\partial \tau}&=-\int_{\Omega} d\bm{x}\;\nabla \phi\cdot\bm{J}, & &\forall \bm{x}\in\Omega\\
\bm{J}\cdot d\bm{\sigma} &= 0, & &\forall \bm{x}\in\partial\Omega
\end{align}
where $\bm{J}$ is defined in Eq.~(\ref{eq:flux}) and $d\bm{\sigma}$ is
the differential boundary-volume element of $\Omega$, with either
finite-element or discontinuous Galerkin methods succeeds in this
feat. While we provide the explicit proof for the full
three-dimensional velocity-space operator, the result holds true also
for the axisymmetric or spherically symmetric cases. We also note that
similar weak discretization of the multispecies collision operator
satisfies the related conservation laws.

\subsection{Time-continuous equation for the degrees-of-freedom}
Choose a \textit{finite}-dimensional vector space $V_h\subset V$ that
is some subset of the space $V$ of all $L^2$-integrable functions in
$\Omega$. Assume that $V_h$ is spanned by the set of functions
$\{\lambda_{\ell}(\bm{x})\}_{\ell}$, and approximate
$(f,\phi)\approx(f_h,\phi_h)$ according to
\begin{align}
f_h(\bm{x},\tau)&=\sum_{\ell}F_{\ell}(\tau)\lambda_{\ell}(\bm{x}),\\
\phi_h(\bm{x})&=\sum_{\ell}\phi_{\ell}\lambda_{\ell}(\bm{x}).
\end{align}
For convinience, denote also the set $F=\{F_{\ell}\}_{\ell}$, i.e.,
the set of degrees-of-freedom for $f_h$. Define the vector- and
tensor-valued functionals
\begin{align}
\bm{K}[F](\tau,\bm{x})&=\sum_{\ell}F_{\ell}(\tau)\bm{K}_{\ell}(\bm{x}),\\
\mathbf{D}[F](\tau,\bm{x})&=\sum_{\ell}F_{\ell}(\tau)\mathbf{D}_{\ell}(\bm{x})
\end{align}
in terms of the vectors $\bm{K}_{\ell}$ and the tensors
$\mathbf{D}_{\ell}$
\begin{align}
\label{eq:fric_diff_coef}
\bm{K}_{\ell}(\bm{x})&=\int_{\Omega}d\bm{\bar{x}}\;\mathbf{U}(\bm{x},\bm{\bar{x}})\cdot\bar{\nabla}\bar{\lambda}_{\ell}(\bm{\bar{x}}),\\
\mathbf{D}_{\ell}(\bm{x})&=\int_{\Omega}d\bm{\bar{x}}\;\mathbf{U}(\bm{x},\bm{\bar{x}})\;\bar{\lambda}_{\ell}(\bm{\bar{x}})
\end{align}
Substitute $f_h$ and $\phi_h$ into Eq.~(\ref{eq:normalized-weak-form})
to obtain the discretized but time-continuous weak formulation
\begin{equation}
\label{eq:fem_weak_nonlinear}
\sum_{ij}\phi_i{\cal M}_{ij}\frac{\partial F_j}{\partial \tau}=\sum_{ij}\phi_i{\cal C}_{ij}[F]\;F_{j},
\end{equation}
where the coefficient matrices are defined
\begin{align}
  {\cal M}_{ij}&=\int_{\Omega} d\bm{x}\; \lambda_i\lambda_j,\\
  {\cal C}_{ij}[F]&=\int_{\Omega}
  d\bm{x}\;\nabla\lambda_i\cdot\left(\bm{K}[F]\;\lambda_j-\mathbf{D}[F]\cdot\nabla\lambda_j\right),
\end{align}
Since the discrete weak form~(\ref{eq:fem_weak_nonlinear}) is to hold
for arbitrary functions $\phi_h\in V_h$, we obtain the following
nonlinear system of ordinary differential equations for the
degrees-of-freedom $F$
\begin{equation}
\label{eq:dofs}
\sum_{j}{\cal M}_{ij}\frac{\partial F_j}{\partial \tau}=\sum_{j}{\cal C}_{ij}[F]\;F_{j},\qquad \forall i.
\end{equation}

\subsection{Discrete conservation laws}
If the vector space $V_h$ is chosen so that the functions
$\phi(\bm{x})=\{1,\bm{x},{\cal E}(\bm{x})\}$ are included in $V_h$
exactly, i.e., $\phi(\bm{x})\equiv\sum_i\phi_i\lambda_i(\bm{x})$, the
weak discretization will automatically satisfy the conservation laws.

Consider the time rate of change of $\phi$-moment of the numerical
distribution function $f_h$. As long as $\phi$ belongs to $V_h$
exactly, we can write
\begin{equation}
\int d\bm{x}\;\phi\;\frac{\partial f_h}{\partial \tau}=\sum_{ij}\phi_i{\cal M}_{ij}\frac{\partial F_j}{\partial t}=\sum_{ij}\phi_i{\cal C}_{ij}[F]\;F_{j}.
\end{equation}
Let us then assume that a quadrature rule is used to approximate
integrals over the domain $\Omega$, with a set of weights $\{w_q\}_q$
and points $\{\bm{\xi}_q\}_{q}$. The vector $\sum_j{\cal
  C}_{ij}[F]F_j$ can then be evaluated as
\begin{multline}
  \sum_j{\cal C}_{ij}[F]F_j=\sum_{j,q}w_qF_j\nabla\lambda_i(\bm{\xi}_q)\cdot\Big(\bm{K}[F](\bm{\xi}_q)\lambda_j(\bm{\xi}_q)\\-\mathbf{D}[F](\bm{\xi}_q)\cdot\nabla\lambda_j(\bm{\xi}_q)\Big).
\end{multline}
The expressions for $\bm{K}[F]$ and $\mathbf{D}[F]$ at the points
$\bm{\xi}_q$ are obtained using the same quadrature rule
\begin{align}
\bm{K}[F](\bm{\xi}_q)&=\sum_{\ell,p}\;w_p\;\mathbf{U}(\bm{\xi}_q,\bm{\xi}_p)\cdot\nabla\lambda_{\ell}(\bm{\xi}_p)F_{\ell},\\
\mathbf{D}[F](\bm{\xi}_q)&=\sum_{\ell,p}\;w_p\;\mathbf{U}(\bm{\xi}_q,\bm{\xi}_p)\lambda_{\ell}(\bm{\xi}_p)F_{\ell},
\end{align}
and, when substituted to the expression for $\sum_j{\cal C}_{ij}[F]F_j$, we find 
\begin{align}
&\sum_j{\cal C}_{ij}[F]F_{j}\nonumber\\
&=\sum_{j\ell,pq}\;w_pw_qF_{\ell}F_j\nabla\lambda_i(\bm{\xi}_q)\cdot\mathbf{U}(\bm{\xi}_q,\bm{\xi}_p)\nonumber\\
&\qquad\cdot\Big(\nabla\lambda_{\ell}(\bm{\xi}_p)\lambda_j(\bm{\xi}_q)-\lambda_{\ell}(\bm{\xi}_p)\nabla\lambda_j(\bm{\xi}_q)\Big).
\end{align}
Since this expression is antisymmetric with respect to changing
$j\leftrightarrow\ell$ and $p\leftrightarrow q$, we obtain
\begin{align}
&\sum_{ij}\phi_i{\cal C}_{ij}[F]F_{j}\nonumber\\
&=\frac{1}{2}\sum_{ij\ell,pq}\;w_pw_qF_{\ell}F_j\phi_i\Big(\nabla\lambda_i(\bm{\xi}_q)-\nabla\lambda_i(\bm{\xi}_p)\Big)\nonumber\\
&\qquad\cdot\mathbf{U}(\bm{\xi}_q,\bm{\xi}_p)\cdot\Big(\nabla\lambda_{\ell}(\bm{\xi}_p)\lambda_j(\bm{\xi}_q)-\lambda_{\ell}(\bm{\xi}_p)\nabla\lambda_j(\bm{\xi}_q)\Big).
\end{align}
The exact conservation laws then follow as in the infinite-dimensional
case since
\begin{equation}
\sum_i\phi_i\Big(\nabla\lambda_i(\bm{\xi}_q)-\nabla\lambda_i(\bm{\xi}_p)\Big)\cdot\mathbf{U}(\bm{\xi}_q,\bm{\xi}_p),
\end{equation}
vanishes identically for $\sum_i\phi_i\lambda_i(\bm{x})\equiv
\{1,\bm{x},{\cal E}(\bm{x})\}$.

Here we wish to note that, in the nonrelativistic limit, the energy
${\cal E}(\bm{x})=x^2$ is a polynomial, and can be exactly expressed
with piecewise polynomials of order 2. Thus a standard finite-element
or discontinuous Galerkin method will have no trouble satisfying the
conservation laws. In the relativistic case, the energy ${\cal
  E}(\bm{x})=\sqrt{1+x^2}$ is, however, not a polynomial and cannot be
presented exactly in terms of piecewise polynomials of any
order. Thus, standard finite-element or discontinuous Galerkin method
will not achieve exact energy conservation in the relativistic case,
although one could still expect the error to converge at the order of
the basis functions. We also point out that in the numerical
integration one has to deal with $\mathbf{U}(\bm{\xi}_q,\bm{\xi}_p)$
which is singular for $q=p$. The total integrand around this
singularity is, however, antisymmetric and thus does not contribute to
the final integral value.

\subsection{A note on discretizing time}
Although our purpose is not to focus on the time discretization -- it
should be chosen consistently with the discretization of the
Vlasov-Maxwell part -- the ordinary differential equation for the
degrees-of-freedom will be nonlinear and stiff due to the presence of
both advective and diffusive components, necessitating implicit time
discretization and iterative methods. Here we comment on the
importance of solving the nonlinear time-discrete equation exactly.

Consider Eq.~(\ref{eq:dofs}) and assume we solve it using implicit
Euler. Denote $F_{\ell}(\tau_k)=F_{\ell}^{(k)}$ so that
\begin{equation}
\frac{\partial F_{\ell}}{\partial \tau}(\tau_k)\approx\frac{F_{\ell}^{(k)}-F_{\ell}^{(k-1)}}{\delta \tau}
\end{equation}
The time discrete equation for the degrees-of-freedom then becomes
\begin{equation}
\sum_{j}{\cal M}_{ij}\left(F_j^{(k)}-F_j^{(k-1)}\right)=\delta\tau\sum_{j}{\cal C}_{ij}[F^{(k)}]\;F^{(k)}_{j},\qquad \forall i.
\end{equation}
Assume then that the iterative method provides us with a solution
vector $\tilde{F}$ that satisfies
\begin{equation}
\sum_{j}{\cal M}_{ij}\left(\tilde{F}_j-F_j^{(k-1)}\right)=\delta\tau\sum_{j}{\cal C}_{ij}[\tilde{F}]\tilde{F}_{j}+\epsilon_i, \quad \forall i
\end{equation}
where $\epsilon=\{\epsilon_i\}_i$ is the residual of the
iteration. For the collisional invariants $\phi(\bm{x})\in
\{1,\bm{x},{\cal E}(\bm{x})\}$, we then have
\begin{equation}
\sum_{ij}\phi_i{\cal M}_{ij}\left(\tilde{F}_j-F_j^{(k-1)}\right)=\sum_i\phi_i\epsilon_i\leq  \lvert\epsilon\rvert_{\infty}\lvert \phi \rvert_{\infty}.
\end{equation}
The exactness of the conservation properties for the discretized
collision operator thus depends only on the accuracy of the nonlinear
solve.

\section{Numerical example}
\label{sec:axial_symmetry}
For demonstration purposes we consider the relaxation of a
nonrelativistic axially symmetric double-Maxwellian distribution
function
\begin{multline}
f(\bm{x})=\frac{1}{2}\left(\pi\sigma^2\right)^{-3/2}\Biggr[ \exp\left(-\frac{r^2+z^2}{\sigma^2}\right)\\+\exp\left(-\frac{r^2+(z-0.5)^2}{\sigma^2}\right)\Biggr],
\end{multline}
using cylindrical coordinates $\bm{x}=(r ,\theta,z)$ that relate to
cartesian coordinates according to $(x,y,z)=(r \cos\theta,r
\sin\theta,z)$. For the computational domain we choose $\Omega=\{(r,z)
\mid 0 \le r \le L, -L\le z \le L\}$ with $L=2$. The parameter
$\sigma=1/\sqrt{20}$ is chosen so that the initial distribution $f$
can be considered negligible at the boundary $\partial\Omega_D=\{(r,z)
\mid z=\pm L \vee r=L\}$.

For the velocity-space discretization, we choose quadratic P2-Lagrange
elements, while time is discretized with the Crank-Nicolson method
using steps of $10^{-3}$ for $\tau$. The resulting nonlinear system is
solved with Newton iteration, using a numerical estimate for the
system Jacobian matrix. Because we do not have an exact linearization
of the Jacobian we only observe linear convergence in the Newton
iteration, with a residual reduction rate of 0.16 for this specific
problem. The P2-mesh is generated with the open-source
\texttt{GMSH}~\cite{gmsh_2009} software with a total of 299
degrees-of-freedom, and the rest of the implementation is carried out
within the \texttt{PETSc}~\cite{petsc-web-page,petsc-user-ref}
framework, using \texttt{PETSc PLEX} for the finite-element operations
and \texttt{PETCs SNES} for the nonlinear solver. The axially
symmetric weak formulation is detailed in the Appendix and the source
code for the test problem, written in \texttt{C}, will be made
available online through \texttt{git}.

\begin{figure*}[!h]
\subfloat[$\tau=0$ \label{sfig:1}]{%
  \includegraphics[width=.3\textwidth]{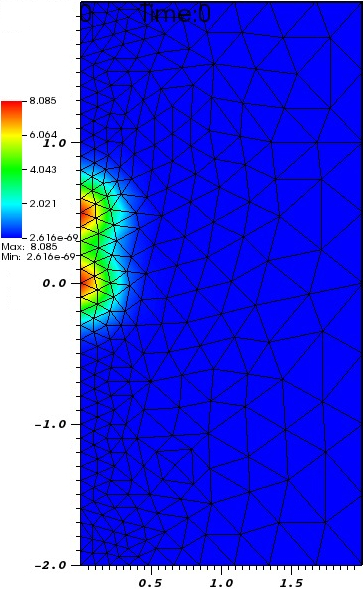}%
}\hfill
\subfloat[$\tau=\delta\tau$ \label{sfig:2}]{%
  \includegraphics[width=.3\textwidth]{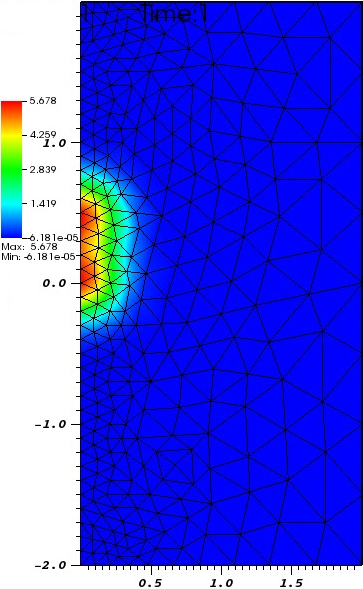}%
}\hfill
\subfloat[$\tau=2\,\delta\tau$ \label{sfig:3}]{%
  \includegraphics[width=.3\textwidth]{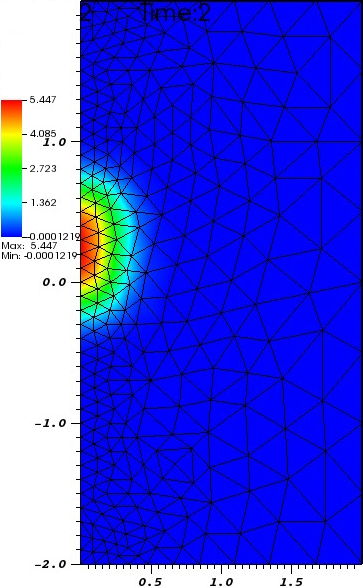}%
}\hfill
\subfloat[$\tau=4\,\delta\tau$ \label{sfig:4}]{%
  \includegraphics[width=.3\textwidth]{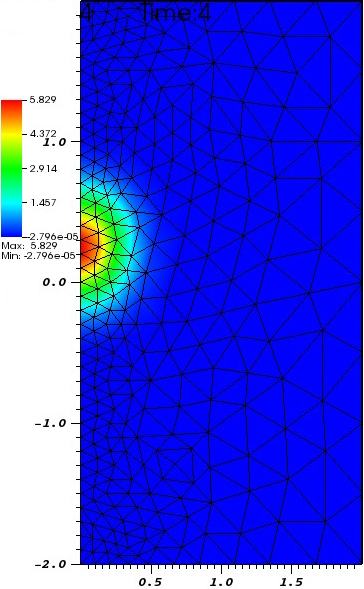}%
}\hfill
\subfloat[$\tau=10\,\delta\tau$ \label{sfig:4}]{%
  \includegraphics[width=.3\textwidth]{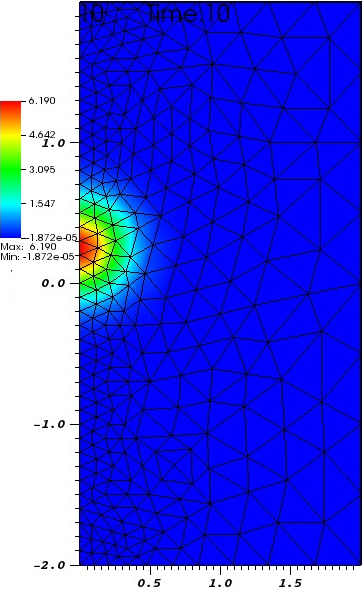}%
}\hfill
\subfloat[$\tau=20\,\delta\tau$ \label{sfig:4}]{%
  \includegraphics[width=.3\textwidth]{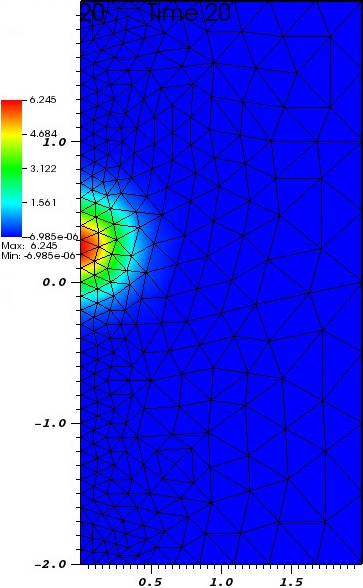}%
}
\caption{Time slices of an initially double Maxwellian distribution function relaxing towards an equilibrium state.}
\label{fig:evolution}
\end{figure*}

The time evolution of the distribution function is illustrated in
Fig.~\ref{fig:evolution}, for six different time instances, while the
evolution of momentum and energy are quantified in
tables~\ref{tbl:momentum} and~\ref{tbl:energy} for different nonlinear
solver tolerances. The double-Maxwellian distribution relaxes towards
an equilibrium state in a qualitatively correct manner and, if the
tolerance for the nonlinear solve is set to machine precision, energy
and momentum are conserved to machine precision. Otherwise the errors
in energy and momentum accumulate through time with a rate that
correlates with the nonlinear solver tolerance.

\begin{table*}[!h]
\caption{\label{tbl:momentum} Momentum conservation, measured with the innerproduct $\sum_{ij}\phi^i{\cal M}_{ij}F_j$ and $\phi=z$ for different nonlinear solver tolerances $\epsilon_{\textrm{tol}}$. The incorrect digits are highlighted with red color.}
\begin{ruledtabular}
\begin{tabular}{llll}
$\#\, \delta\tau$ & $\epsilon_{\textrm{tol}}=$\textrm{9.0E-01}  & $\epsilon_{\textrm{tol}}=$\textrm{1.0E-06} & $\epsilon_{\textrm{tol}}=$\textrm{1.0E-14} \\
\hline
0 & 3.97797664845241E-02              & 3.97797664845241E-02              & 3.97797664845248E-02 \\
1 & 3.9779{\color{red}1615978734}E-02 & 3.97797664{\color{red}781777}E-02 & 3.97797664845248e-02 \\
2 & 3.977{\color{red}88912692161}E-02 & 3.97797664{\color{red}743008}E-02 & 3.9779766484524{\color{red}7}E-02 \\
4 & 3.977{\color{red}86187222186}E-02 & 3.97797664{\color{red}643968}E-02 & 3.9779766484524{\color{red}7}E-02 \\
10& 3.977{\color{red}81761433900}E-02 & 3.97797664{\color{red}423772}E-02 & 3.9779766484524{\color{red}7}E-02 \\
20& 3.977{\color{red}75989653156}E-02 & 3.9779766{\color{red}3011934}E-02 & 3.9779766484524{\color{red}7}E-02 \\
\end{tabular}
\end{ruledtabular}
\caption{\label{tbl:energy} Energy conservation, measured with the innerproduct $\sum_{ij}\phi^i{\cal M}_{ij}F_j$ and $\phi=r^2+z^2$ for different nonlinear solver tolerances $\epsilon_{\textrm{tol}}$. The incorrect digits are highlighted with red color.}
\begin{ruledtabular}
\begin{tabular}{llll}
$\#\, \delta\tau$ & $\epsilon_{\textrm{tol}}=$\textrm{9.0E-01}  & $\epsilon_{\textrm{tol}}=$\textrm{1.0E-06} & $\epsilon_{\textrm{tol}}=$\textrm{1.0E-14} \\
\hline
0 & 3.17986788742740E-02              & 3.17986788742740E-02              & 3.17986788742740E-02 \\
1 & 3.17{\color{red}606259814932}E-02 & 3.179867{\color{red}90469298}E-02 & 3.17986788742740E-02 \\
2 & 3.17{\color{red}718385841557}E-02 & 3.179867{\color{red}94847238}E-02 & 3.1798678874274{\color{red}1}E-02 \\
4 & 3.1{\color{red}8140462853466}E-02 & 3.17986{\color{red}808661484}E-02 & 3.1798678874274{\color{red}1}E-02 \\
10& 3.1{\color{red}8822521328414}E-02 & 3.17986{\color{red}851426220}E-02 & 3.1798678874274{\color{red}1}E-02 \\ 
20& 3.1{\color{red}9077804661782}E-02 & 3.17986{\color{red}927385047}E-02 & 3.1798678874274{\color{red}2}E-02 \\
\end{tabular}
\end{ruledtabular}
\end{table*}

\section{Summary}
\label{sec:summary}
We have presented an algorithm for conservative discretization of the
nonlinear Landau collision integral. We have provided both algebraic
and numerical proof for achieving exact numerical conservation laws
using either discontinuous Galerkin or standard finite-element
method. Our method is not constrained by details of the
discretization, admitting the use of structurized as well as
unstructurized meshes. We have also argued that in the relativistic
case, a polynomial basis of any order is not able to guarantee exact
conservation of energy while density and momentum would be conserved
even with linear basis functions. Future study will investigate the
embedding of our discrete Landau operator to the Vlasov-Maxwell system
using either the concept of Lagrange-d'Alembert principle or extended
Lagrangians~\cite{Kraus_Maj_2015}. Another future study will focus on
performance demonstrations.

\begin{acknowledgments}
  The work presented here was carried out as part of the Simulation
  Center for Runaway Electron Mitigation and Avoidance efforts. The
  work by EH was supported by U.S. Department of Energy Contract
  No. DE-AC02-09-CH11466 and the work by MFA was supported by the
  Director, Office of Science, and Office of Advanced Scientific
  Computing Research of the U.S. Department of Energy under Contract
  No. D-AC02-05CH11231. The authors appreciate the help provided by
  Prof. Matthew Knepley who patiently answered all questions related
  to the use of \texttt{PETSc}.
\end{acknowledgments}

\appendix*
\section{Axially symmetric weak formulation}
Using cylindrical coordinates $\bm{x}=(r ,\theta,z)$, that relate to
cartesian coordinates according to $(x,y,z)=(r \cos\theta,r
\sin\theta,z)$, and assuming axially symmetric vector space $V$, i.e.,
$\partial f/\partial \theta = 0$ and $\partial \phi/\partial
\theta=0$, the weak formulation can be written
\begin{equation}
2\pi\int_{\Omega} dr  dz\; r \; \phi\; \partial_{\tau}f = \int_{\Omega} dr  dz\; r \; \partial_{\alpha}\phi\;\left(K^{\alpha}f- D^{\alpha\beta}\partial_{\beta}f\right)
\end{equation}
where the friction and diffusion coefficients are
\begin{align}
\label{eq:Ka}
K^{\alpha}&=\int_{\Omega}d\bar{r }d\bar{z}\;\bar{r }\;U^{\alpha\bar{\beta}}\partial_{\bar{\beta}}\bar{f}\\
\label{eq:Dab}
D^{\alpha\beta}&=\int_{\Omega}d\bar{r }d\bar{z}\;\bar{r }\;U^{\alpha\beta}\bar{f},
\end{align}
and the coefficients $U^{\alpha\beta}$ and $U^{\alpha\bar{\beta}}$ are defined
\begin{align}
U^{\alpha\beta}=\int_0^{2\pi}\int_0^{2\pi}d\theta d\bar{\theta}\; \nabla x^{\alpha}\cdot\mathbf{U}\cdot\nabla x^{\beta} \\
U^{\alpha\bar{\beta}}=\int_0^{2\pi}\int_0^{2\pi}d\theta d\bar{\theta}\; \nabla x^{\alpha}\cdot\mathbf{U}\cdot\bar{\nabla}\bar{x}^{\beta}.
\end{align}
The expressions $\nabla x^{\alpha}$ are the contravariant basis
vectors for the curvilinear coordinate system.

For the nonrelativistic case, the angular integrals of $\nabla
x^{\alpha}\cdot\mathbf{U}\cdot\nabla x^{\beta}$ are easily
computed. Defining a parameter
\begin{equation}
s(r ,z,\bar{r },\bar{z})=\frac{2r \bar{r }}{r^2+\bar{r}^2+(z-\bar{z})^2},
\end{equation}
the exact expressions are
\begin{align}
U^{rr}&= 4\pi\left(\frac{s}{2r \bar{r }}\right)^{3/2}(\bar{r}^2{\cal I}_1+(z-\bar{z})^2{\cal I}_2)\\
U^{r z}&= 4\pi\left(\frac{s}{2r\bar{r}}\right)^{3/2}(\bar{z}-z)(r{\cal I}_2-\bar{r}{\cal I}_3)\\
U^{z r}&=U^{r z}\\
U^{z z}&= 4\pi\left(\frac{s}{2r\bar{r}}\right)^{3/2}((r^2+\bar{r}^2){\cal I}_2-2r\bar{r}{\cal I}_3)\\
U^{r\bar{r}}&=4\pi\left(\frac{s}{2r\bar{r}}\right)^{3/2}((z-\bar{z})^2{\cal I}_3+r\bar{r}{\cal I}_1)\\
U^{r\bar{z}}&=U^{r z}\\
U^{z \bar{r}}&=4\pi\left(\frac{s}{2r\bar{r}}\right)^{3/2}(\bar{z}-z)(r{\cal I}_3-\bar{r}{\cal I}_2)\\
U^{z \bar{z}}&=U^{zz}
\end{align}
where the integrals ${\cal I}(s)$ are defined
\begin{align}
{\cal I}_1(s)&=\int_{-1}^1(1-x^2)^{1/2}(1-sx)^{-3/2}dx\\
{\cal I}_2(s)&=\int_{-1}^1(1-x^2)^{-1/2}(1-sx)^{-3/2}dx\\
{\cal I}_3(s)&=\int_{-1}^1x(1-x^2)^{-1/2}(1-sx)^{-3/2}dx
\end{align}
and can be expressed in terms of the complete elliptic integrals
$E[s]$ and $K[s]$ according to
\begin{align}
{\cal I}_1(s)&=\frac{4}{s^2\sqrt{1+s}}\left(K\left[\frac{2s}{1+s}\right]-(1+s)E\left[\frac{2s}{1+s}\right]\right),\\
{\cal I}_2(s)&=\frac{2}{(1-s)\sqrt{1+s}}E\left[\frac{2s}{1+s}\right],\\
{\cal I}_3(s)&=\frac{2}{(1-s)s\sqrt{1+s}}\left(E\left[\frac{2s}{1+s}\right]-(1-s)K\left[\frac{2s}{1+s}\right]\right)
\end{align}
A similar computation of the axisymmetric coefficients was
demonstrated in~\cite{Yoon_et_al_2014}.

\bibliographystyle{unsrt}
\bibliography{bibfile}

\end{document}